\documentclass[aps,prc,twocolumn,superscriptaddress,showpacs]{revtex4}

\usepackage{graphicx}

\begin{document}

\title{ 
Isoscaling in Peripheral Nuclear Collisions 
around the Fermi Energy and a Signal of Chemical Separation 
from its Excitation Energy Dependence 
      }

\author{M. Veselsky}
\email{fyzimarv@savba.sk}
\affiliation{Cyclotron Institute, Texas A\&M University, College Station, TX 77843}
\affiliation{Institute of Physics of the Slovak Academy of Sciences, Bratislava, Slovakia}
\author{G. A. Souliotis}
\affiliation{Cyclotron Institute, Texas A\&M University, College Station, TX 77843}
\author{S. J. Yennello}
\affiliation{Cyclotron Institute, Texas A\&M University, College Station, TX 77843}

\date{\today}

\nopagebreak

\begin{abstract}
The isoscaling is investigated using the fragment yield data 
from fully reconstructed quasi-projectiles observed in peripheral collisions 
of $^{28}$Si with  $^{124,112}$Sn at projectile energies 
30 and 50 MeV/nucleon. The excitation energy dependence of the isoscaling 
parameter $\beta^{\prime}$ is observed which is 
independent of beam energy. For a given quasi-projectile 
produced in reactions with different targets no isoscaling is observed. 
The isoscaling thus reflects the level of N/Z-equilibration 
in reactions with different targets represented by the initial 
quasi-projectile samples. The excitation energy dependence of 
the isoscaling parameter $\beta^{\prime}$, corrected for the trivial 1/T 
temperature dependence, does not follow the trend of the homogeneous system 
above 4 MeV/nucleon thus possibly signaling the onset of separation 
into isospin asymmetric dilute and isospin symmetric dense phase.

\end{abstract}



\maketitle


The isotopic composition of nuclear reaction products provides 
important information on the reaction dynamics and the possible occurrence of 
a phase transition in the asymmetric nuclear matter \cite{Serot,Andrej}, 
which is supposed to lead to separation into a symmetric dense 
phase and asymmetric dilute phase. It has been discussed in literature 
\cite{Instab,Baran,Colonna,Spinod} to what extent such a phase transition 
is generated by fluctuations of density or concentration, typically 
suggesting a coupling of different instability modes. 
The N/Z ( neutron to proton ratio ) 
degree of freedom and its equilibration was studied experimentally 
in detailed measurements of the isotopic distributions of emitted fragments 
\cite{SJY1,Johnston,Ram,RLSiSn,MVSiSn,Rami,Xu}. Isotopically resolved data 
in the region Z=2--8 have revealed systematic trends, which however 
are substantially affected by the decay of the excited primary fragments. 
It has recently been shown  \cite{Tsang1} that the effect 
of sequential decay of primary fragments can be overcome by comparing 
the yields of fragments from two similar reactions. For statistical fragment 
production in two reactions with different isospin asymmetry, but at the same 
temperature, the ratio R$_{21}$(N,Z) of the yields of a given fragment (N,Z) 
exhibits an exponential dependence on N and Z. 
This scaling behavior is termed isoscaling 
\cite{Tsang1} and has been observed in a variety of reactions under 
the  conditions of statistical emission and  equal temperature
\cite{Tsang2,Tsang3,IsoBotvina,GSHRIso}. 

In this study, we present an isoscaling analysis 
of a recent fragment yield data \cite{RLSiSn,MVSiSn,MVIsoDist,MVIsoTemp} 
obtained by charged-particle calorimetry of hot thermally equilibrated 
quasi-projectiles from the reactions $^{28}$Si+$^{124,112}$Sn at 
30 and 50 MeV/nucleon. 
The angular coverage and granularity of the forward-angle multidetector 
array FAUST \cite{Faust} allowed full reconstruction of the hot 
quasiprojectiles from mid-peripheral binary collisions \cite{MVSiSn}. 
Excellent description of the fragment observables 
was obtained using the model of deep-inelastic transfer ( DIT ) \cite{Tassan} 
for the early stage of collisions and the statistical multifragmentation 
model ( SMM ) \cite{SMM} for de-excitation. 
The large N/Z range of quasi-projectiles allowed 
unique studies of the dependence of fragment observables 
on quasi-projectile N/Z. An increasingly inhomogeneous N/Z-distribution 
between light charged particles and heavier fragments was 
observed with increasing isospin asymmetry \cite{MVIsoDist}. 
An exponential scaling of the isobaric ratio Y($^{3}$H)/Y($^{3}$He) 
with the quasi-projectile N/Z was observed and the temperature was 
extracted from the logarithmic slope \cite{MVIsoTemp}. 
In the present work we examine the isoscaling phenomenon 
in unique conditions of highly-selective data. Typically, 
the investigations of isoscaling  focused 
on yields of light fragments with Z=2-8 originating from de-excitation of 
massive hot systems produced using reactions of mass symmetric projectile and 
target at intermediate energies \cite{Tsang1,Tsang2,Tsang3} or by reactions 
of high-energy light particle with massive target nucleus 
\cite{IsoBotvina}. In a recent article \cite{GSHRIso}, 
we report the investigations of isoscaling using the heavy residue 
data from the reactions of 25 MeV/nucleon $^{86}$Kr projectiles 
with  $^{124}$Sn,$^{112}$Sn and $^{64}$Ni, $^{58}$Ni targets. 
Here we present the investigation of the isoscaling phenomenon 
on the full sample of fragments emitted by the hot thermally 
equilibrated quasi-projectiles with mass A=20-30.




A thermally equilibrated system undergoing statistical decay can be, 
within grand-canonical approach, characterized by a yield of fragments 
with neutron and proton numbers N and Z 
\cite{DoubleRat,Randrup}: 

\begin{equation}
        Y(N,Z) = F(N,Z)\exp[B(N,Z)/T]\exp(N \mu_{n}/T + Z \mu_{p}/T)
\label{eqn1}
\end{equation}

where $F(N,Z)$ represents contribution due to the secondary decay from  
particle stable and  unstable states to the ground state; $\mu_{n}$ 
and $\mu_{p}$ are the free neutron and proton chemical potentials; $B(N,Z)$ is 
the ground state binding energy of the fragment, and $T$ is the 
temperature.
\par
The ratio of the isotope yields from two different systems,  having similar 
excitation energies and similar masses, but differing only
in N/Z, cancels out the effect of secondary decay and provides information 
about the excited primary fragments \cite{Tsang1}. Within the grand-canonical 
approximation ( Eq. \ref{eqn1} ), the ratio $Y_{2}(N,Z)/Y_{1}(N,Z)$ assumes 
the form 

\begin{equation}
     R_{21}(N,Z) = Y_{2}(N,Z)/Y_{1}(N,Z) = C \exp(\alpha N  + \beta Z)
\end {equation}       

with  $\alpha$ = $\Delta \mu_{n}$/T 
and $\beta$ = $\Delta \mu_{p}$/T, with $\Delta \mu_{n}$ and $\Delta \mu_{p}$ 
being the differences in the free neutron  and proton chemical potentials 
of the fragmenting systems. C is an  overall normalization constant.
Alternatively, as introduced by Botvina et al. \cite{IsoBotvina}, 
the isoscaling can be expressed as

\begin{equation}
     R_{21}(N,Z) = Y_{2}(N,Z)/Y_{1}(N,Z) 
     = C \exp(\alpha^{\prime} A  + \beta^{\prime} (N-Z) )
\end {equation}       

thus introducing the scaling parameters which can be related to the isoscalar 
and isovector components of free nucleon chemical potential since 
$\alpha^{\prime}$ = $\Delta (\mu_{n}+\mu_{p})$/2T 
and $\beta^{\prime}$ = $\Delta (\mu_{n}-\mu_{p})$/2T. 
In this variant of isoscaling, the mass dependence is rather weak and the 
information is concentrated in the isoscaling parameter $\beta^{\prime}$. 


    \begin{figure}[h]                                        

   \includegraphics[width=0.44\textwidth, height=0.33\textheight ]{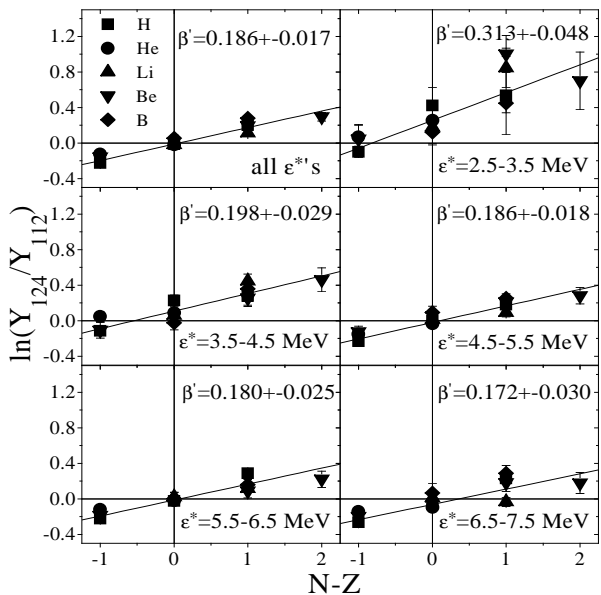}

    \caption{
    The isoscaling plots from the reactions of $^{28}$Si+$^{124,112}$Sn 
    at 50 MeV/nucleon for the full set of isotopically resolved quasi-projectiles 
    and for five bins of quasi-projectile excitation energy per nucleon. 
           }
    \label{Iso50}
    \end{figure}

\begin{figure}[h]                                        

   \includegraphics[width=0.44\textwidth, height=0.33\textheight ]{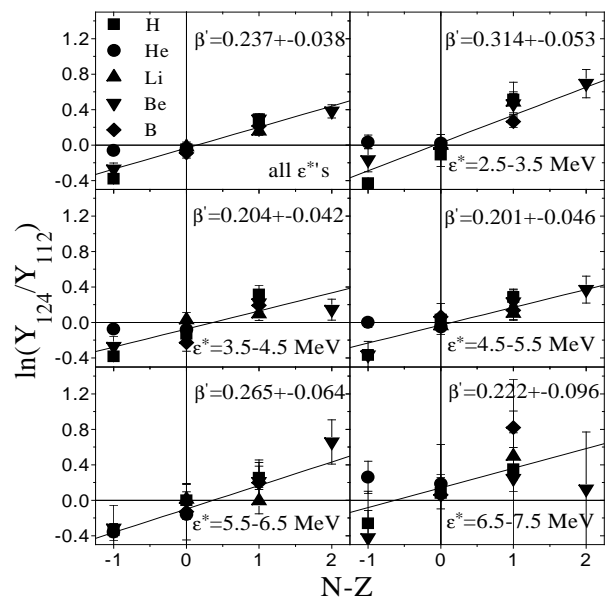}

    \caption{
    As in Fig. \ref{Iso50} but for 30 MeV/nucleon.
           }
    \label{Iso30}
    \end{figure}

In Figs. \ref{Iso50} and \ref{Iso30}, we present the isoscaling data 
from statistical decay of hot quasiprojectiles from the reactions 
$^{28}$Si+$^{124,112}$Sn at projectile energy 50 and 30 MeV/nucleon, 
respectively. Observed charged 
particles with Z$\leq$5 have been isotopically resolved and the 
total observed charge was close to the charge of the projectile ( Z=12-15 )
\cite{RLSiSn,MVSiSn}.  The observed 
fragment data provide full information ( with the exception of emitted 
neutrons ) on the de-excitation of thermally equilibrated hot quasi-projectiles 
with known mass ( A=20-30 ), charge, velocity and excitation energy
\cite{MVSiSn}. In Figs. \ref{Iso50} and \ref{Iso30}, the isoscaling plots are 
presented not only for the full data but also for five 
bins of excitation energy. The isoscaling 
occurs at both projectile energies, both for the inclusive data and 
for the sub-sets with different excitation energies. 
The slope depends on the excitation energy in a similar way 
as the slope of the dependence of isobaric ratio Y($^{3}$H)/Y($^{3}$He) 
on the quasi-projectile N/Z observed in \cite{MVIsoTemp}. 
On the other hand, the isoscaling parameters for individual excitation energy 
bins do not depend on the projectile energy and the difference of the slopes 
of the inclusive data is given by the different excitation 
energy distributions of hot quasi-projectiles at two projectile energies. 


\begin{figure}[h]                                        

   \includegraphics[width=0.335\textwidth, height=0.275\textheight ]{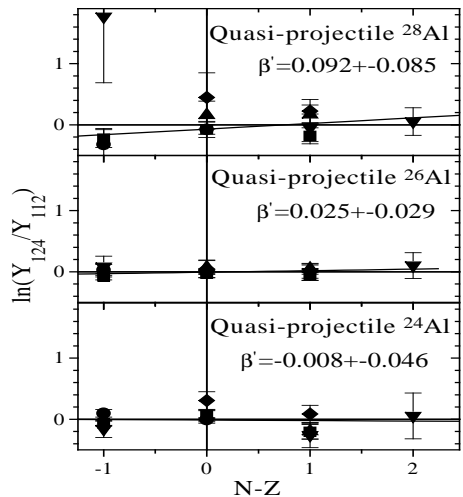}

    \caption{
    The isoscaling plots from the reactions of $^{28}$Si+$^{124,112}$Sn 
    at 50 MeV/nucleon for three reconstructed quasi-projectiles. 
    Symbols as in Figs. \ref{Iso50},\ref{Iso30}. 
           }
    \label{NoIso}
    \end{figure}

In Fig. \ref{NoIso} are presented isoscaling plots for three 
selected quasiprojectiles $^{24,26,28}$Al ( representing the central part 
of quasiprojectile mass and charge distribution ) in the reactions 
of $^{28}$Si+$^{124,112}$Sn targets at projectile energy 50 MeV/nucleon. 
Isoscaling behavior is not observed which means that 
no memory of the entrance channel interaction leading to production 
of a given quasi-projectile remains and the isoscaling analysis 
is not sensitive to the statistical fluctuations in de-excitation. 
The slight hint of isoscaling in the case of $^{28}$Al may be 
due to low statistics or due to larger uncertainty 
in the number of unobserved neutrons of the reconstructed neutron-rich 
quasiprojectile $^{28}$Al caused by more intense neutron emission 
than for proton-rich isotopes $^{24,26}$Al.  
A behavior similar in appearance to Fig. \ref{NoIso} 
was reported by Tsang et al. \cite{Tsang2} using the yields of fragments  
with Z=3-8 from the reaction of $^{16}$O with  $^{197}$Au, $^{208}$Pb targets 
at projectile energy 20 MeV/nucleon \cite{OAuPb}, measured at 40$^{\circ}$. 
There, the absence of isoscaling behavior suggests that the 
sample of original prefragments is independent of the target isospin 
asymmetry, what for such inclusive data suggests a geometric 
reaction scenario such as breakup or abrasion and the observed fragments 
are most probably the spectator parts of the original projectile, ejected 
to a large laboratory angle by the recoil in the projectile breakup. 


\begin{figure}[h]                                        

   \includegraphics[width=0.35\textwidth, height=0.25\textheight ]{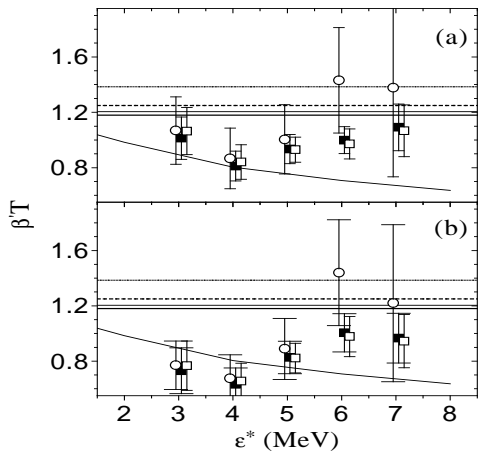}

    \caption{
    The dependence of observable $\beta^{\prime} T$ on relative 
    excitation energy of the quasiprojectile. 
    Open squares and circles show experimental data obtained 
    using the double isotope ratio (a) and Y($^{3}$H)/Y($^{3}$He) slope (b) 
    temperature at 50 and 30 MeV/nucleon, respectively. Full squares 
    represent combined data from both energies. 
    Full and dashed horizontal 
    lines show estimate of $\beta^{\prime} T$ obtained when assuming
    $\mu_{p,n} \approx -S_{p,n}$ at 50 and 30 MeV/nucleon, respectively. 
    Solid curve represents the expected trend for homogeneous system. 
    For details see text. 
           }
    \label{IsoTmp}
    \end{figure}

As suggested by Botvina et al. \cite{IsoBotvina}, the isoscaling parameter 
trends are similar to the observed double isotope ratio 
temperature. The temperature for the current data was studied in detail 
in \cite{MVIsoTemp} and we use the results of that work for comparison 
with the isoscaling parameters. In Fig. \ref{IsoTmp} 
we present the observable $\beta^{\prime} T$, canceling 
out the trivial 1/T-dependence of the isoscaling 
parameter. In Fig. \ref{IsoTmp}a we use the double isotope ratio 
temperature \cite{DoubleRat} from the isotopic ratios 
Y($^{2}$H)/Y($^{3}$H) and Y($^{3}$He)/Y($^{4}$He) 
and in Fig. \ref{IsoTmp}b we use the temperature 
obtained from the slope of the dependence of 
Y($^{3}$H)/Y($^{3}$He) on the quasi-projectile N/Z ( mirror nucleus 
temperature ) \cite{MVIsoTemp}. 
Grossly, the $\beta^{\prime} T$ should reflect the level of N/Z equilibration 
\cite{GSHRIso}, which is known for the current data \cite{MVSiSn} where 
full N/Z-equilibration was not reached. 
The horizontal lines represent the estimate of the zero temperature values 
of $\beta^{\prime} T$ of reconstructed quasi-projectiles 
using the known N/Z equilibration \cite{MVSiSn}. 
We estimate the zero temperature proton and neutron 
chemical potentials by the proton and neutron 
separation energies ( $\mu_{p,n} \approx -S_{p,n}$ ). 
Similar approximation was used to extract 
the mirror nucleus temperature \cite{MVIsoTemp}. 
The zero temperature estimates of $\beta^{\prime} T$ for 
all excitation energy bins are consistent with horizontal lines 
which is a result of the selection of quasiprojectiles with Z=12-15. 
Any experimental deviations 
from the zero temperature values of $\beta^{\prime} T$ can be understood as 
a non-trivial dependence representing the details of de-excitation. 
The thick and thin lines represent the $\beta^{\prime} T$ estimates 
with and without the correction for neutron emission 
from backtracing of DIT/SMM simulations \cite{MVSiSn}. 
The dashed/full lines represent the projectile 
energies 30 and 50 MeV/nucleon, respectively. The open circles/squares 
represent the experimental data at 30 and 50 MeV/nucleon while the 
full squares represent the combined data from both energies. 
The experimental dependences of $\beta^{\prime} T$ on excitation energies 
indicate an initial decrease at low excitation energies 
( consistent with expansion of the homogeneous excited source ), 
further they exhibit a turning-point at 4 MeV/nucleon 
followed by increase toward the zero temperature value at 6-7 MeV/nucleon. 
The trend is similar at both projectile energies independent 
of the method of temperature extraction and quality of the statistical sample. 
The existence of such a turning-point can be understood in the context 
of work \cite{MVIsoTemp}, where the viability of the approximation  
$\mu_{p,n} \approx -S_{p,n}$ at high temperatures 
was supported by a possible counterbalance of two effects, namely expansion 
of hot nucleus ( leading to a decrease of the 
absolute values of chemical potentials compared to the zero temperature values 
as indicated by a solid curve ) and chemical separation into an isospin 
symmetric heavy fraction and an isospin asymmetric nucleon gas, leading 
to stronger dependence of $\mu_{n} - \mu_{p}$ of the nucleon gas 
( dilute phase ) on N/Z of the system \cite{Serot,Andrej} thus resulting 
in higher sensitivity of the observed yield ratios. 
Such a conclusion is supported by 
the inhomogeneous isospin distribution observed for the current 
data \cite{MVIsoDist} where the isospin asymmetry of light charged particles 
exhibits strong sensitivity on N/Z of the quasiprojectile while the isospin 
asymmetry of heavier fragments is rather insensitive.
The estimate of $\beta^{\prime} T$ for 
homogeneous system was obtained assuming $\rho^{2/3}$-dependence 
and using the estimate of free volume 
( eq. 52 of ref. \cite{SMM} ) for each excitation energy bin 
from the successful DIT/SMM simulations.  Since the isoscaling parameter 
depends directly on the free nucleon chemical potentials, 
the turning-point at 4 MeV/nucleon can be understood as a signal of the onset 
of chemical separation which reverts the decrease of the free nucleon 
chemical potential consistent with expansion of the homogeneous system. 
This is also consistent with the recent theoretical study 
of thermodynamical properties of finite nuclei of Sanzhur et al. \cite{Andrej2} 
where the hot nucleus ( at a given initial excitation energy ) 
can expand up to the turning point 
where the central part begins to contract again while the increasing 
number of nucleons form the dilute component surrounding the dense 
central part. The values of the temperature characterizing the 
turning point in the central density for a given excitation energy reproduce 
well the caloric curve \cite{MVIsoTemp} for the data investigated here. 
The existence of the turning-point explains the difference 
of caloric curves ( around 4 MeV/nucleon ) obtained in \cite{MVSiSn} 
using double isotope ratio and mirror nucleus temperatures, former free of 
assumptions on chemical potentials ( the same applying to Fig. \ref{IsoTmp}a ) 
while latter using the approximation $\mu_{p,n} \approx -S_{p,n}$ ( 
thus artificially lowering the values of $\beta^{\prime} T$ around 
the turning-point ). When estimating the density of the homogeneous system at 
the turning point ( assuming $\rho^{2/3}$-dependence of $\beta^{\prime} T$ ), 
it is consistent with theoretical estimates of the position and shape 
of the spinodal contour, typically at almost constant total density 0.6$\rho_0$ in the wide range 
of asymmetries \cite{Spinod}. Thus, inside the spinodal region the homogeneous 
nuclear medium is quickly replaced by the inhomogeneous system. 


In summary, the isoscaling is investigated using the fragment yield data 
from fully reconstructed quasi-projectiles observed in peripheral collisions 
of $^{28}$Si with  $^{124,112}$Sn at projectile energies 
30 and 50 MeV/nucleon. The excitation energy dependence of the isoscaling 
parameter $\beta^{\prime}$ is observed which is 
independent of the beam energy. For a given reconstructed quasi-projectile 
produced in reactions with different targets no isoscaling is observed. 
Thus, the magnitude of the isoscaling is determined 
by the difference of initial quasi-projectile samples produced 
in reactions with different targets leading to different level 
of N/Z-equilibration. The excitation energy 
dependence of the observable $\beta^{\prime} T$, corrected 
for the 1/T temperature dependence, exhibits a turning-point at 4 MeV/nucleon 
which can be interpreted as a signal of the onset of separation 
into isospin asymmetric dilute and isospin symmetric dense phase. The 
initial trend of expansion of the homogeneous system is reverted 
by the transition into an inhomogeneous system with two phases 
having different proton and neutron concentrations. A novel 
experimental information on the equation of state of the asymmetric nuclear 
matter is thus obtained.


\par

We are thankful to L. Tassan-Got for his DIT code and to A. Botvina 
for his SMM code. This work was supported in part by the Robert A. 
Welch Foundation through grant No. A-1266, the Department of Energy
through grant No. DE-FG03-93ER40773, and the Slovak Scientific Grant 
Agency through grant VEGA-2/1132/21. 


\bibliography{sisniso5.bib}

\begin{thebibliography}{27}
\expandafter\ifx\csname natexlab\endcsname\relax\def\natexlab#1{#1}\fi
\expandafter\ifx\csname bibnamefont\endcsname\relax
  \def\bibnamefont#1{#1}\fi
\expandafter\ifx\csname bibfnamefont\endcsname\relax
  \def\bibfnamefont#1{#1}\fi
\expandafter\ifx\csname citenamefont\endcsname\relax
  \def\citenamefont#1{#1}\fi
\expandafter\ifx\csname url\endcsname\relax
  \def\url#1{\texttt{#1}}\fi
\expandafter\ifx\csname urlprefix\endcsname\relax\def\urlprefix{URL }\fi
\providecommand{\bibinfo}[2]{#2}
\providecommand{\eprint}[2][]{\url{#2}}

\bibitem[{\citenamefont{Muller and Serot}(1995)}]{Serot}
\bibinfo{author}{\bibfnamefont{H.}~\bibnamefont{Muller}} \bibnamefont{and}
  \bibinfo{author}{\bibfnamefont{B.~D.} \bibnamefont{Serot}},
  \bibinfo{journal}{Phys. Rev. C} \textbf{\bibinfo{volume}{52}},
  \bibinfo{pages}{2072} (\bibinfo{year}{1995}).

\bibitem[{\citenamefont{Kolomietz et~al.}(2001)}]{Andrej}
\bibinfo{author}{\bibfnamefont{V.~M.} \bibnamefont{Kolomietz}}
  \bibnamefont{et~al.}, \bibinfo{journal}{Phys. Rev. C}
  \textbf{\bibinfo{volume}{64}}, \bibinfo{pages}{24315} (\bibinfo{year}{2001}).

\bibitem[{\citenamefont{Li and Ko}(1997)}]{Instab}
\bibinfo{author}{\bibfnamefont{B.-A.} \bibnamefont{Li}} \bibnamefont{and}
  \bibinfo{author}{\bibfnamefont{C.}~\bibnamefont{Ko}}, \bibinfo{journal}{Nucl.
  Phys. A} \textbf{\bibinfo{volume}{618}}, \bibinfo{pages}{498}
  (\bibinfo{year}{1997}).

\bibitem[{\citenamefont{Baran et~al.}(2001)}]{Baran}
\bibinfo{author}{\bibfnamefont{V.}~\bibnamefont{Baran}} \bibnamefont{et~al.},
  \bibinfo{journal}{Phys. Rev. Lett} \textbf{\bibinfo{volume}{86}},
  \bibinfo{pages}{4492} (\bibinfo{year}{2001}).

\bibitem[{\citenamefont{Colonna et~al.}(2002)}]{Colonna}
\bibinfo{author}{\bibfnamefont{M.}~\bibnamefont{Colonna}} \bibnamefont{et~al.},
  \bibinfo{journal}{Phys. Rev. Lett} \textbf{\bibinfo{volume}{88}},
  \bibinfo{pages}{122701} (\bibinfo{year}{2002}).

\bibitem[{\citenamefont{Marqueron and Chomaz}(2003)}]{Spinod}
\bibinfo{author}{\bibfnamefont{J.}~\bibnamefont{Marqueron}} \bibnamefont{and}
  \bibinfo{author}{\bibfnamefont{P.}~\bibnamefont{Chomaz}},
  \bibinfo{journal}{Phys. Rev. C} \textbf{\bibinfo{volume}{67}},
  \bibinfo{pages}{41602(R)} (\bibinfo{year}{2003}).

\bibitem[{\citenamefont{Yennello et~al.}(1994)}]{SJY1}
\bibinfo{author}{\bibfnamefont{S.~J.} \bibnamefont{Yennello}}
  \bibnamefont{et~al.}, \bibinfo{journal}{Phys. Lett. B}
  \textbf{\bibinfo{volume}{321}}, \bibinfo{pages}{15} (\bibinfo{year}{1994}).

\bibitem[{\citenamefont{Johnston et~al.}(1996)}]{Johnston}
\bibinfo{author}{\bibfnamefont{H.}~\bibnamefont{Johnston}}
  \bibnamefont{et~al.}, \bibinfo{journal}{Phys. Lett. B}
  \textbf{\bibinfo{volume}{371}}, \bibinfo{pages}{186} (\bibinfo{year}{1996}).

\bibitem[{\citenamefont{Ramakrishnan et~al.}(1998)}]{Ram}
\bibinfo{author}{\bibfnamefont{E.}~\bibnamefont{Ramakrishnan}}
  \bibnamefont{et~al.}, \bibinfo{journal}{Phys. Rev. C}
  \textbf{\bibinfo{volume}{57}}, \bibinfo{pages}{1803} (\bibinfo{year}{1998}).

\bibitem[{\citenamefont{Laforest et~al.}(1999)}]{RLSiSn}
\bibinfo{author}{\bibfnamefont{R.}~\bibnamefont{Laforest}}
  \bibnamefont{et~al.}, \bibinfo{journal}{Phys. Rev. C}
  \textbf{\bibinfo{volume}{59}}, \bibinfo{pages}{2567} (\bibinfo{year}{1999}).

\bibitem[{\citenamefont{Veselsky et~al.}(2000{\natexlab{a}})}]{MVSiSn}
\bibinfo{author}{\bibfnamefont{M.}~\bibnamefont{Veselsky}}
  \bibnamefont{et~al.}, \bibinfo{journal}{Phys. Rev. C}
  \textbf{\bibinfo{volume}{62}}, \bibinfo{pages}{064613}
  (\bibinfo{year}{2000}{\natexlab{a}}).

\bibitem[{\citenamefont{Rami et~al.}(2000)}]{Rami}
\bibinfo{author}{\bibfnamefont{F.}~\bibnamefont{Rami}} \bibnamefont{et~al.},
  \bibinfo{journal}{Phys. Rev. Lett.} \textbf{\bibinfo{volume}{84}},
  \bibinfo{pages}{1120} (\bibinfo{year}{2000}).

\bibitem[{\citenamefont{Xu et~al.}(2000)}]{Xu}
\bibinfo{author}{\bibfnamefont{H.~S.} \bibnamefont{Xu}} \bibnamefont{et~al.},
  \bibinfo{journal}{Phys. Rev. Lett.} \textbf{\bibinfo{volume}{85}},
  \bibinfo{pages}{716} (\bibinfo{year}{2000}).

\bibitem[{\citenamefont{Tsang et~al.}(2001{\natexlab{a}})}]{Tsang1}
\bibinfo{author}{\bibfnamefont{M.~B.} \bibnamefont{Tsang}}
  \bibnamefont{et~al.}, \bibinfo{journal}{Phys. Rev. Lett.}
  \textbf{\bibinfo{volume}{86}}, \bibinfo{pages}{5023}
  (\bibinfo{year}{2001}{\natexlab{a}}).

\bibitem[{\citenamefont{Tsang et~al.}(2001{\natexlab{b}})}]{Tsang2}
\bibinfo{author}{\bibfnamefont{M.~B.} \bibnamefont{Tsang}}
  \bibnamefont{et~al.}, \bibinfo{journal}{Phys. Rev. C}
  \textbf{\bibinfo{volume}{64}}, \bibinfo{pages}{041603}
  (\bibinfo{year}{2001}{\natexlab{b}}).

\bibitem[{\citenamefont{Tsang et~al.}(2001{\natexlab{c}})}]{Tsang3}
\bibinfo{author}{\bibfnamefont{M.~B.} \bibnamefont{Tsang}}
  \bibnamefont{et~al.}, \bibinfo{journal}{Phys. Rev. C}
  \textbf{\bibinfo{volume}{64}}, \bibinfo{pages}{054615}
  (\bibinfo{year}{2001}{\natexlab{c}}).

\bibitem[{\citenamefont{Botvina et~al.}(2002)\citenamefont{Botvina, Lozhkin,
  and Trautmann}}]{IsoBotvina}
\bibinfo{author}{\bibfnamefont{A.~S.} \bibnamefont{Botvina}},
  \bibinfo{author}{\bibfnamefont{O.~V.} \bibnamefont{Lozhkin}},
  \bibnamefont{and}
  \bibinfo{author}{\bibfnamefont{W.}~\bibnamefont{Trautmann}},
  \bibinfo{journal}{Phys. Rev. C} \textbf{\bibinfo{volume}{65}},
  \bibinfo{pages}{044610} (\bibinfo{year}{2002}).

\bibitem[{\citenamefont{Souliotis et~al.}(2003)}]{GSHRIso}
\bibinfo{author}{\bibfnamefont{G.~A.} \bibnamefont{Souliotis}}
  \bibnamefont{et~al.}, \bibinfo{journal}{Phys. Rev. C}
  \textbf{\bibinfo{volume}{68}}, \bibinfo{pages}{24605} (\bibinfo{year}{2003}).

\bibitem[{\citenamefont{Veselsky et~al.}(2000{\natexlab{b}})}]{MVIsoDist}
\bibinfo{author}{\bibfnamefont{M.}~\bibnamefont{Veselsky}}
  \bibnamefont{et~al.}, \bibinfo{journal}{Phys. Rev. C}
  \textbf{\bibinfo{volume}{62}}, \bibinfo{pages}{041605}
  (\bibinfo{year}{2000}{\natexlab{b}}).

\bibitem[{\citenamefont{Veselsky et~al.}(2001)}]{MVIsoTemp}
\bibinfo{author}{\bibfnamefont{M.}~\bibnamefont{Veselsky}}
  \bibnamefont{et~al.}, \bibinfo{journal}{Phys. Lett. B}
  \textbf{\bibinfo{volume}{497}}, \bibinfo{pages}{1} (\bibinfo{year}{2001}).

\bibitem[{\citenamefont{Gimeno-Nogues et~al.}(1997)}]{Faust}
\bibinfo{author}{\bibfnamefont{F.}~\bibnamefont{Gimeno-Nogues}}
  \bibnamefont{et~al.}, \bibinfo{journal}{Nucl. Instr. and Meth. A}
  \textbf{\bibinfo{volume}{399}}, \bibinfo{pages}{94} (\bibinfo{year}{1997}).

\bibitem[{\citenamefont{Tassan-Got and Stefan}(1991)}]{Tassan}
\bibinfo{author}{\bibfnamefont{L.}~\bibnamefont{Tassan-Got}} \bibnamefont{and}
  \bibinfo{author}{\bibfnamefont{C.}~\bibnamefont{Stefan}},
  \bibinfo{journal}{Nucl. Phys. A} \textbf{\bibinfo{volume}{524}},
  \bibinfo{pages}{121} (\bibinfo{year}{1991}).

\bibitem[{\citenamefont{Bondorf et~al.}(1995)}]{SMM}
\bibinfo{author}{\bibfnamefont{J.~P.} \bibnamefont{Bondorf}}
  \bibnamefont{et~al.}, \bibinfo{journal}{Phys. Rep.}
  \textbf{\bibinfo{volume}{257}}, \bibinfo{pages}{133} (\bibinfo{year}{1995}).

\bibitem[{\citenamefont{Albergo et~al.}(1985)}]{DoubleRat}
\bibinfo{author}{\bibfnamefont{S.}~\bibnamefont{Albergo}} \bibnamefont{et~al.},
  \bibinfo{journal}{Nuovo Cimento A} \textbf{\bibinfo{volume}{89}},
  \bibinfo{pages}{1} (\bibinfo{year}{1985}).

\bibitem[{\citenamefont{Randrup and Koonin}(1981)}]{Randrup}
\bibinfo{author}{\bibfnamefont{J.}~\bibnamefont{Randrup}} \bibnamefont{and}
  \bibinfo{author}{\bibfnamefont{S.~E.} \bibnamefont{Koonin}},
  \bibinfo{journal}{Nucl. Phys. A} \textbf{\bibinfo{volume}{356}},
  \bibinfo{pages}{223} (\bibinfo{year}{1981}).

\bibitem[{\citenamefont{Gelbke et~al.}(1978)}]{OAuPb}
\bibinfo{author}{\bibfnamefont{C.~K.} \bibnamefont{Gelbke}}
  \bibnamefont{et~al.}, \bibinfo{journal}{Phys. Rep.}
  \textbf{\bibinfo{volume}{42}}, \bibinfo{pages}{311} (\bibinfo{year}{1978}).

\bibitem[{\citenamefont{Sanzhur}()}]{Andrej2}
\bibinfo{author}{\bibfnamefont{A.~I.} \bibnamefont{Sanzhur}},
  \bibinfo{howpublished}{private communication}.

\end{thebibliography}



\end{document}